\documentclass[5p]{elsarticle}
\usepackage{graphicx}
\usepackage{dcolumn}
\usepackage{bm}
\usepackage{hyperref}
\usepackage[T1]{fontenc}
\usepackage{pslatex}
\usepackage{textcomp} 
\usepackage{color}
\usepackage{lmodern} 

\hypersetup{
        colorlinks = true,
        pdftitle = {}, 
        pdfauthor = {},
        pdfkeywords = {},
        linkcolor = blue,
        citecolor = blue,
        filecolor = black,
        urlcolor = magenta
}

\begin{document}
\begin{sloppypar}

\title{ \textit{Ab initio} study of pressure-induced phase transition,
band gaps and X-ray photoemission valence band spectra of YVO$_4$}

\author[ifm]{M. Werwi\'nski\corref{cor1}}
\ead{werwinski@ifmpan.poznan.pl}

\author[ifm]{J. Kaczkowski}

\author[ifm]{P. Le\'{s}niak}

\author[ifm]{W.L. Malinowski}

\author[ifm]{A. Szajek}

\author[uam]{A. Szczeszak}

\author[uam]{S. Lis}

\address[ifm]{Institute of Molecular Physics, Polish Academy of Sciences,\\ ul. M. Smoluchowskiego 17, 60-179 Pozna\'n, Poland}

\address[uam]{Adam Mickiewicz University, Faculty of Chemistry, Department of Rare Earths, ul. Umultowska 89b, 61-614 Pozna\'{n}, Poland}

\cortext[cor1]{Corresponding author} 

\date{\today}

\newcommand{\yvo}{YVO$_4$}

\begin{abstract}
High-pressure induced structural transition from zircon-type phase into scheelite-type phase in \yvo{} is studied using \textit{ab initio} calculations.
Several structures with compressed volumes are evaluated, where for every considered volume the c/a ratio and atomic positions are optimised.
The transition pressure and transition volume change are calculated.
The reports on \yvo{} electronic structure and electronic band gap behaviour are followed by results of X-Ray photoemission spectra XPS calculations.
Most of our theoretical predictions are compared with experimental results taken from literature.
\end{abstract}



\maketitle

\section{Introduction}

The Nd-doped crystals of a vanadates family, such as Nd:\yvo{}, have recently been studied as materials for construction of efficient solid-state lasers~\cite{Sato03,Ryba-Romanowski03}.
In turn colloidal Eu-doped \yvo{} is considered as a material for displays applications~\cite{Huignard02}.
Many other representatives of the orthovanadates family are also promising candidates for cathodoluminescent materials, thermophosphorus sensors, or scintillators~\cite{Mullica96}.
It was shown experimentally that a pressure induced structural transition from zircon-type phase into scheelite-type phase in \yvo{}~\cite{Wang04} is ensuing by a significant drops in volume~\cite{Wang04} and in absorption edge~\cite{Panchal11}.
Around 25 GPa the second phase transition takes place into the monoclinic M-fergusonite structure~\cite{Manjon10}.
The zircon-scheelite transition is a first order non-reversible transition and it is common for alkaline-earth vanadates~\cite{Errandonea15}.

In this work the structural transition in \yvo{} is explored using \textit{ab initio} method, despite the fact that the electronic structures of several orthovanadates, including \yvo{}, have been recently calculated~\cite{Panchal11,Manjon10,Huang12,Huang13}.
The strength and novelty of our approach is due to the facts that all the structural parameters of \yvo{} are obtained \textit{ab initio}, which makes our theoretical models parameter free and moreover it is the first time when a full-potential method~\cite{Koepernik99} was used to solve an electronic structure of \yvo, in opposition to a pseudo-potential method applied in the all previous calculations~\cite{Panchal11,Manjon10,Huang12,Huang13}. 
It has been shown, e.g.~\cite{Edstrom15}, that the full-potential may have significant advantages above the approximations.
The optical properties of \yvo{}, drawing the biggest attention, are also addressed by calculating pressure dependencies of electronic band gaps and X-Ray photoemission spectra.

\begin{figure}[ht]
\centering
\includegraphics[trim = 0 0 0 20,clip,width=0.8\columnwidth]{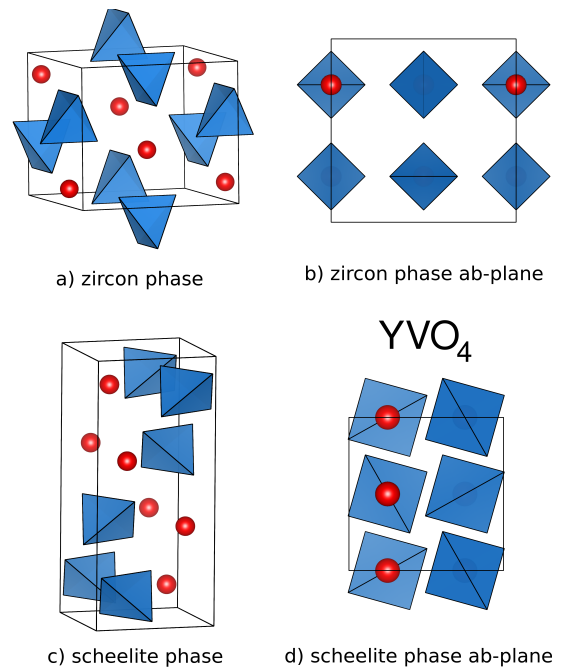}
\caption{\label{fig:structs} The zircon-type and scheelite-type crystal structures of \yvo{} in 
isometric and ab-planes projections. Circles indicate Y atoms. Tetrahedrons are centered with V atoms and tetrahedrons vertices represent O atoms.
}
\end{figure}

\begin{table*}[!ht]
\caption{\label{tab:crystal_data} 
The calculated optimised structural parameters and bulk modulus B$_0$ for zircon phase (sg. I41/amd) and scheelite phase (sg. I41/a) of \yvo{} at zero pressure.
The Wyckoff positions for zircon phase are Y(0, 3/4, 1/8) and V(0, 1/4, 3/8) and for scheelite phase Y(0, 1/4, 5/8) and V(0, 1/4, 1/8)~\cite{Wang04}.
The optimized oxygen Wyckoff positions x, y, and z are shown.
The theoretical results are compared with available literature data, both experimental and theoretical with the same GGA approach.\\
}
\centering
\small
\begin{tabular}{l|lllllll}
\hline
phase                			& $a$ [\AA{}]	& $c$ [\AA{}]&V$_{cell}$ [\AA{}$^3$]& B$_0$ [GPa]& $x(O)$ 	& $y(O)$ 	& $z(O)$ \\
\hline
zircon exp.\cite{Wang04}		&   7.1224(1) 	& 6.29130(12)   & 319.15(1)	& 130(3)& 0  		& 0.9325(5) 	& 0.8010(4)\\
zircon GGA	     			&   7.229   	& 6.333     	& 330.96	& 126  	& 0    		& 0.93339	& 0.79984 \\
zircon GGA \cite{Manjon10}&   7.218	& 6.334		& 330.00	& 123.3	& & & \\
zircon GGA \cite{Huang12}	&   7.18	& 6.314		& 325.50	& 120.0 & & & \\
\hline
scheelite exp.\cite{Wang04}		& 5.0320(3) 	& 11.233(1)     & 284.48(3)	& 138(9)& 0.2454(11)	& 0.6090(11)	& 0.5441(5) \\
scheelite GGA 		  		&   5.077   	& 11.374     	& 293.207	& 142	& 0.24495      	& 0.60624	& 0.54507 \\
scheelite GGA\cite{Manjon10}&  5.080	& 11.348	& 292.85 	& 144.4 & & & \\
scheelite GGA\cite{Huang13}&   5.06	& 11.298	& 289.27	& 126.8, 136.7 & & &  \\
\hline
\end{tabular}
\end{table*}

\begin{figure}[ht]
\centering
\includegraphics[trim = 80 20 0 60,clip,height=\columnwidth,angle=270]{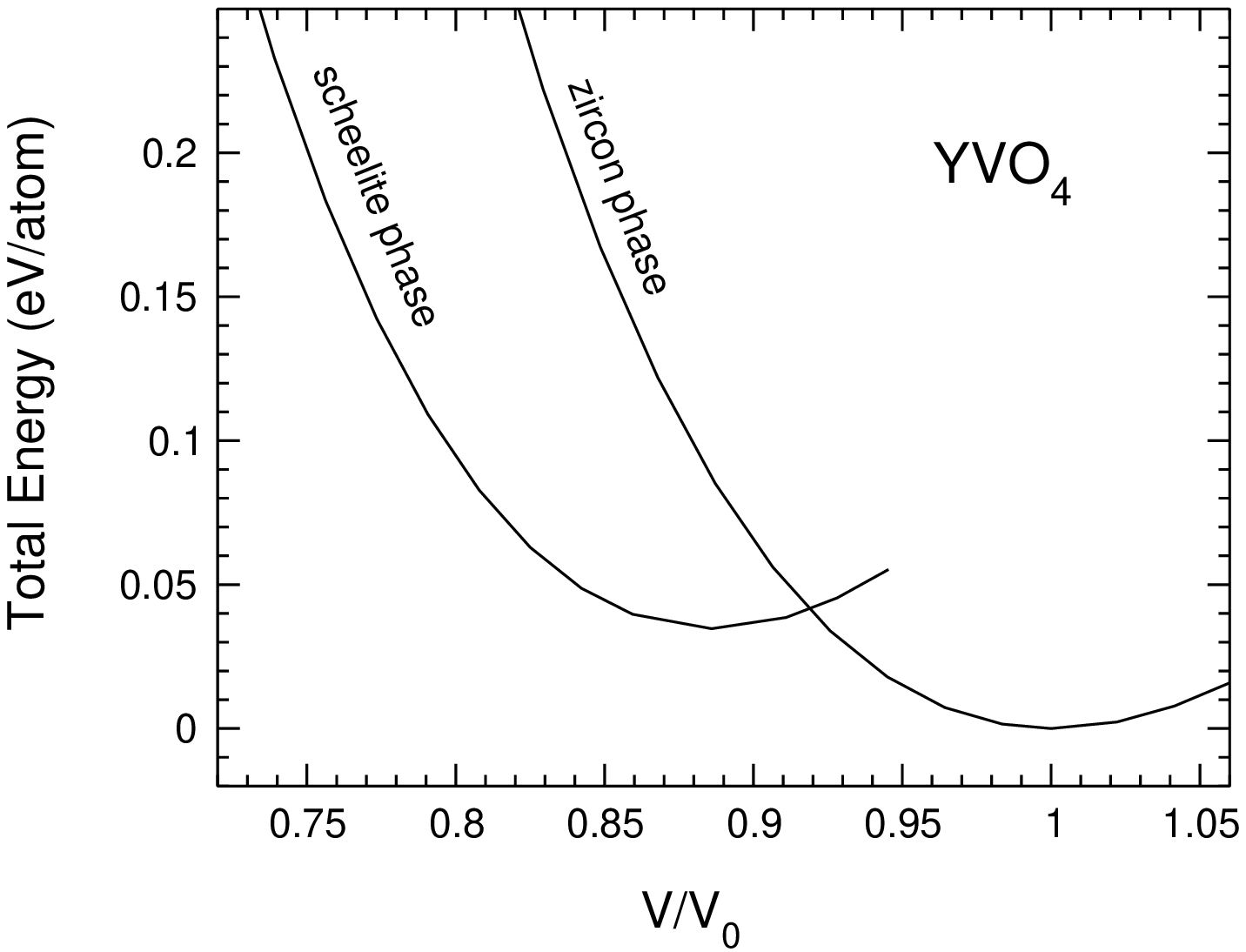}
\caption{\label{fig:E_V} Calculated total energy vs. volume for the zircon and scheelite phases of \yvo{}. V$_0$ is the calculated zero-pressure volume for the zircon phase. The energy of the zircon phase is arbitrary chosen to be zero at zero pressure. A crossing point at V/V$_0$~=~0.919 indicates volume 304.198~\AA{}$^3$/unit cell.
}
\end{figure}

\begin{figure}[ht]
\centering
\includegraphics[trim = 80 20 0 20,clip,height=\columnwidth,angle=270]{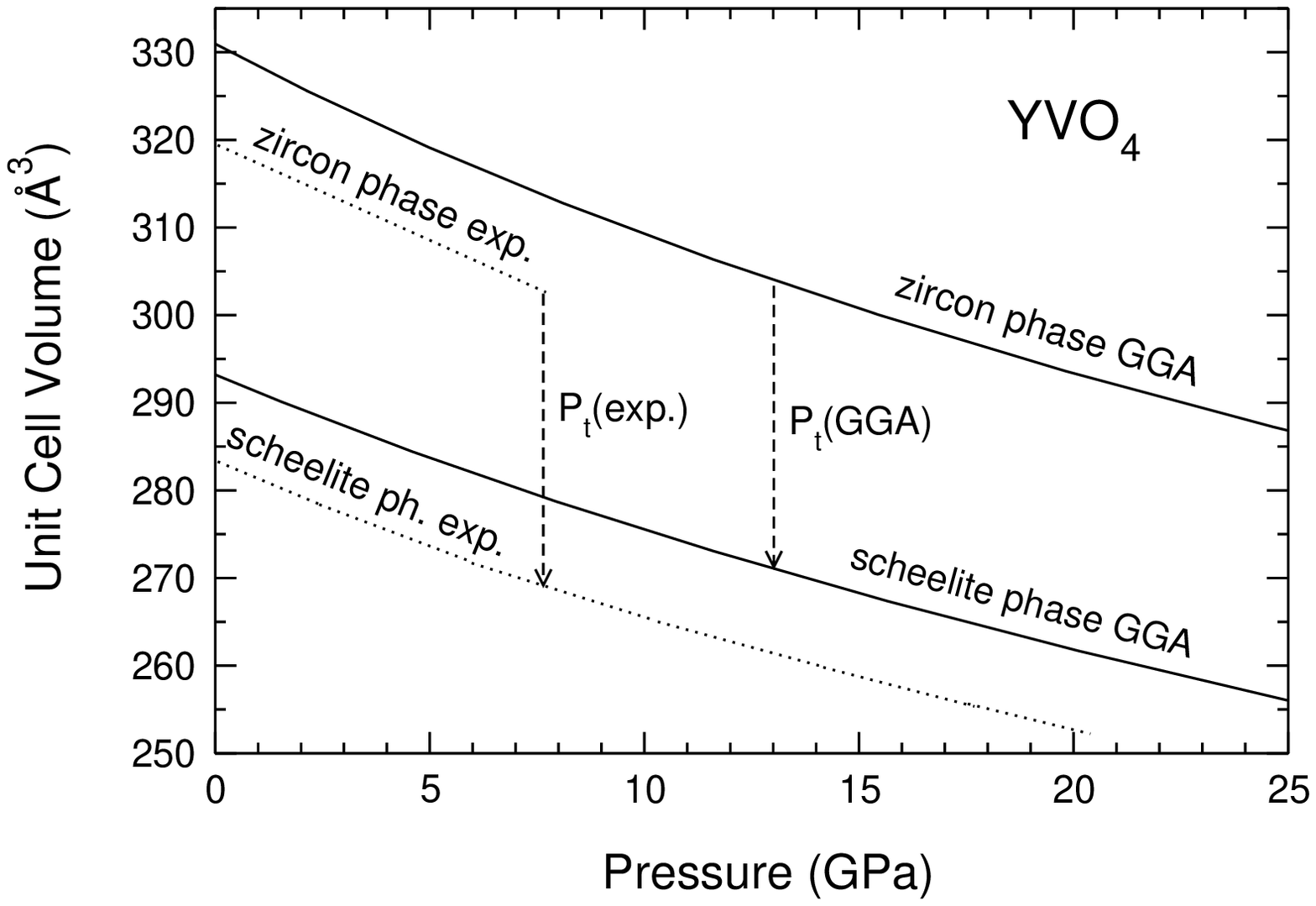}
\caption{\label{fig:V_P} Calculated volume vs. pressure for the zircon and scheelite phases of \yvo{} as obtained by fitting the third-order Birch-Murnaghan equation of state~\cite{Birch47} to the energy-volume data. The solid lines represent calculated values, while the dashed lines represent available experimental data~\cite{Wang04}.}
\end{figure}

\begin{figure}[ht]
\centering
\includegraphics[trim = 50 15 40 80,clip,width=\columnwidth]{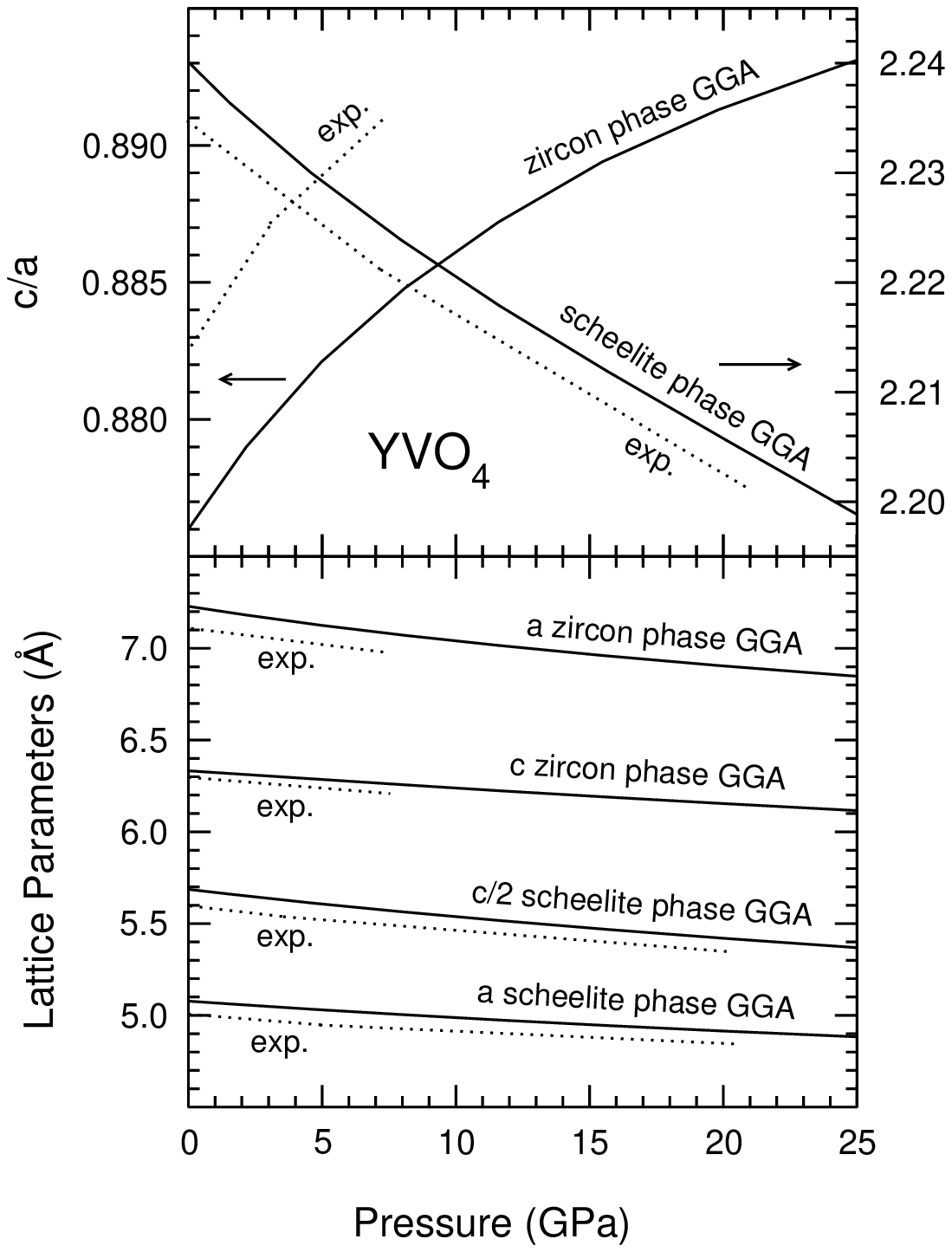}
\caption{\label{fig:lat_param_P} c/a ratio and lattice parameters as a function of pressure for the zircon and scheelite phases of \yvo{}.
The solid lines represent calculated values, while the dashed lines represent available experimental data~\cite{Wang04}.
}
\end{figure}

\begin{figure}[ht]
\centering
\includegraphics[trim = 0 0 0 0,clip,height=\columnwidth,angle=270]{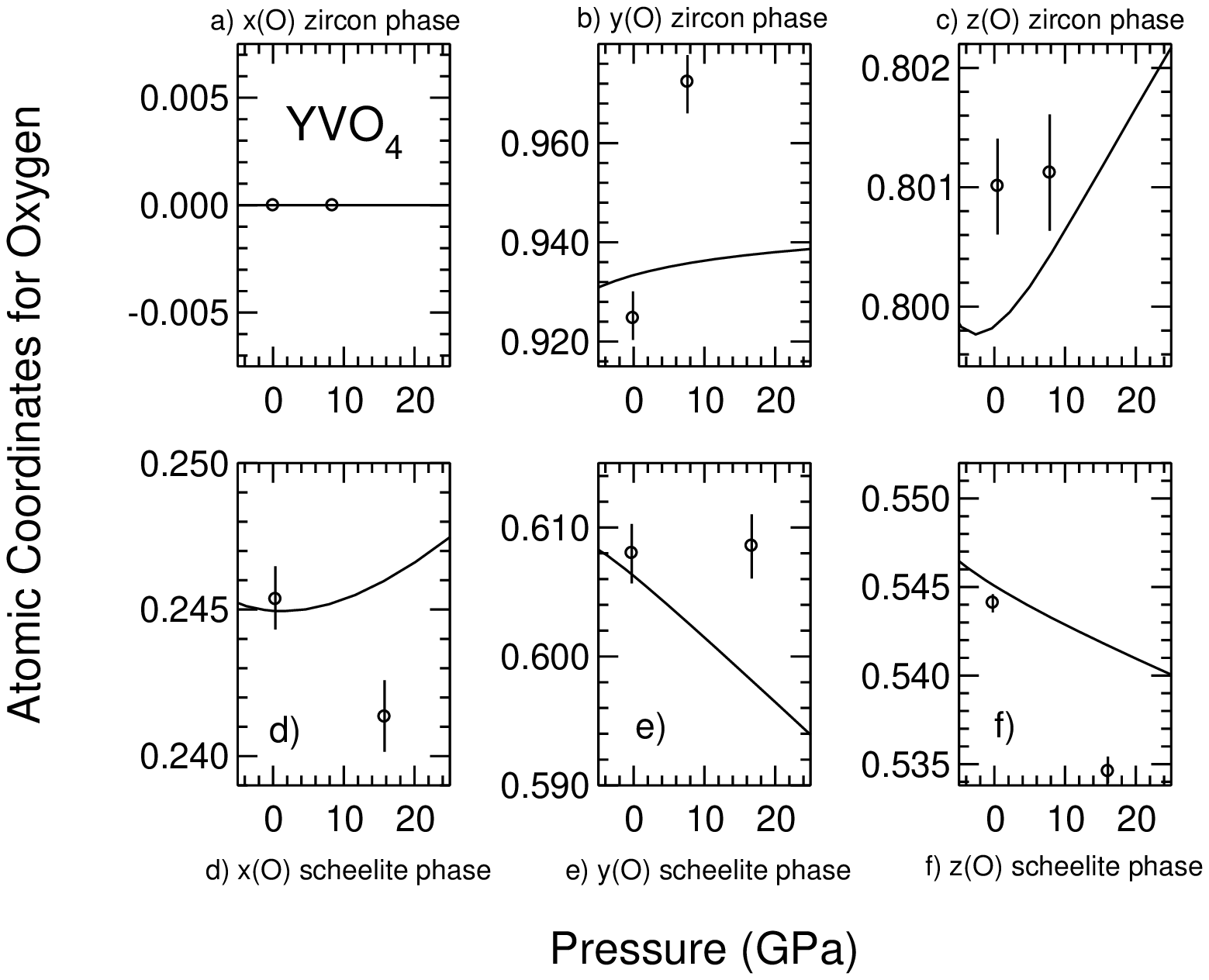}
\caption{\label{fig:positions_P} Atomic positions of oxygen atoms as a function of pressure for the zircon and scheelite phases of \yvo{}. The solid lines represent calculated values. The open circles with error bars represent available experimental data~\cite{Wang04}.
}
\end{figure}

\begin{figure}[ht]
\centering
\includegraphics[trim = 80 0 0 60,clip,height=\columnwidth,angle=270]{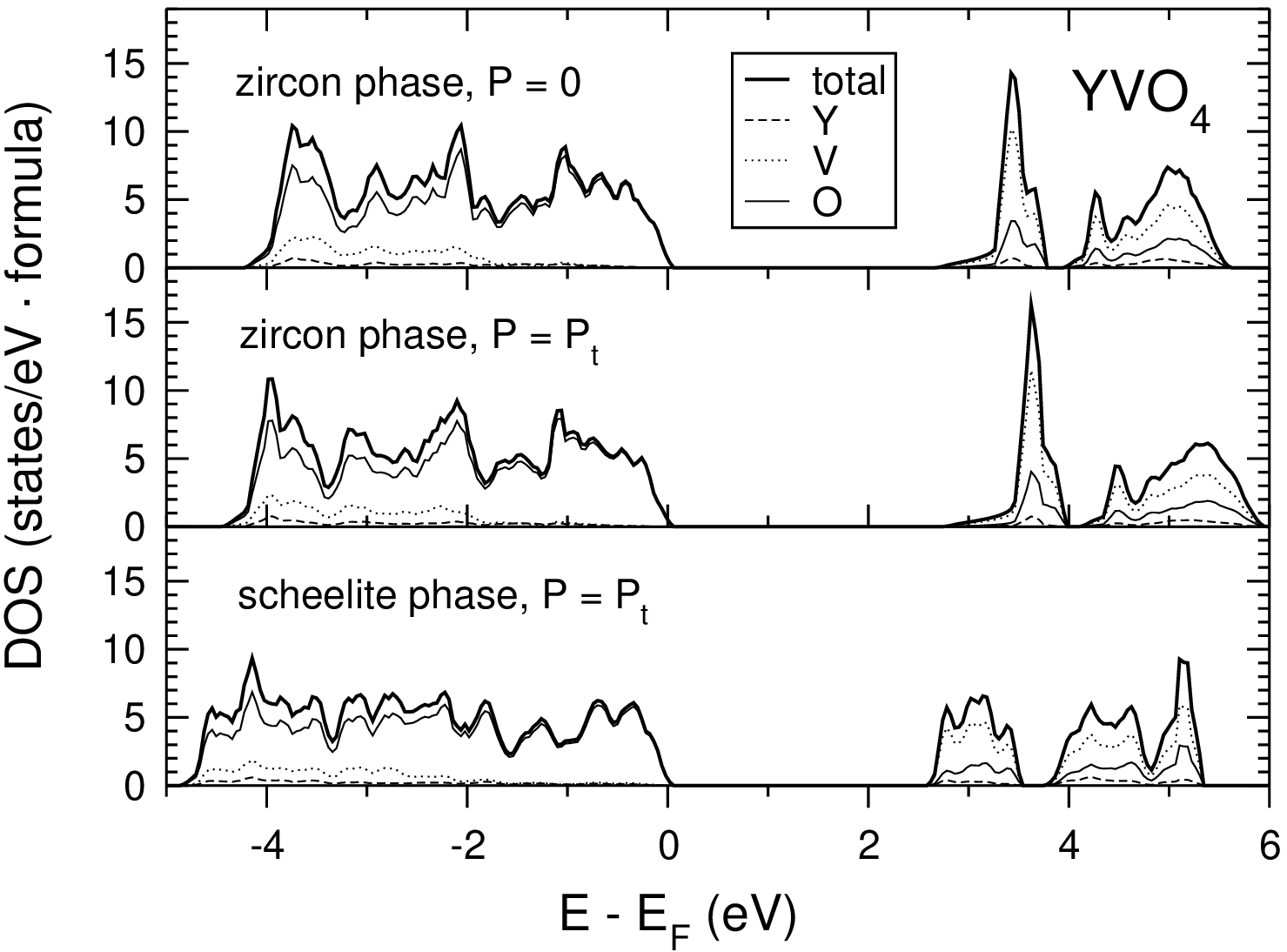}
\caption{\label{fig:dos} Calculated partial and total densities of states (DOS) for the zero pressure and calculated transition pressure P$_t$(GGA)~=~12.95~GPa for the zircon and scheelite-type phases of \yvo{}. Results obtained by the GGA-FPLO14 method. Energy scale shifted to Fermi energy.}
\end{figure}

\begin{figure}[ht]
\centering
\includegraphics[trim = 80 10 0 30,clip,height=\columnwidth,angle=270]{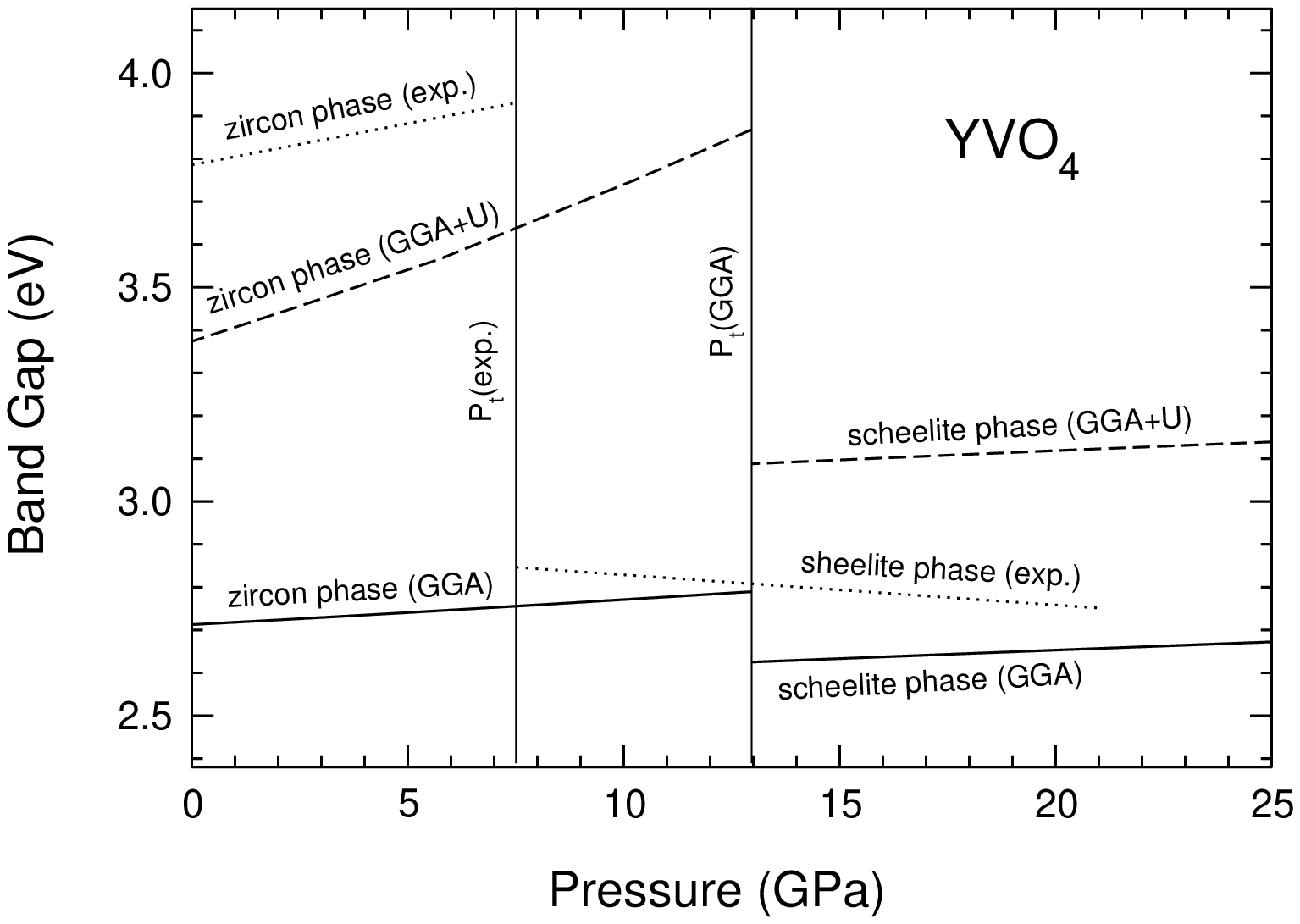}
\caption{\label{fig:gap} Calculated band gap as function of pressure for the zircon and scheelite phases of \yvo{}. 
Results obtained with the GGA and GGA+U (U(V)~=~3.5~eV, U(O)~=~4.5~eV) FPLO14 method are compared with available experimental data~\cite{Cai13,Kruczek05}.
}
\end{figure}

\begin{figure}[ht]
\centering
\includegraphics[trim = 150 20 00 30,clip,height=\columnwidth,angle=270]{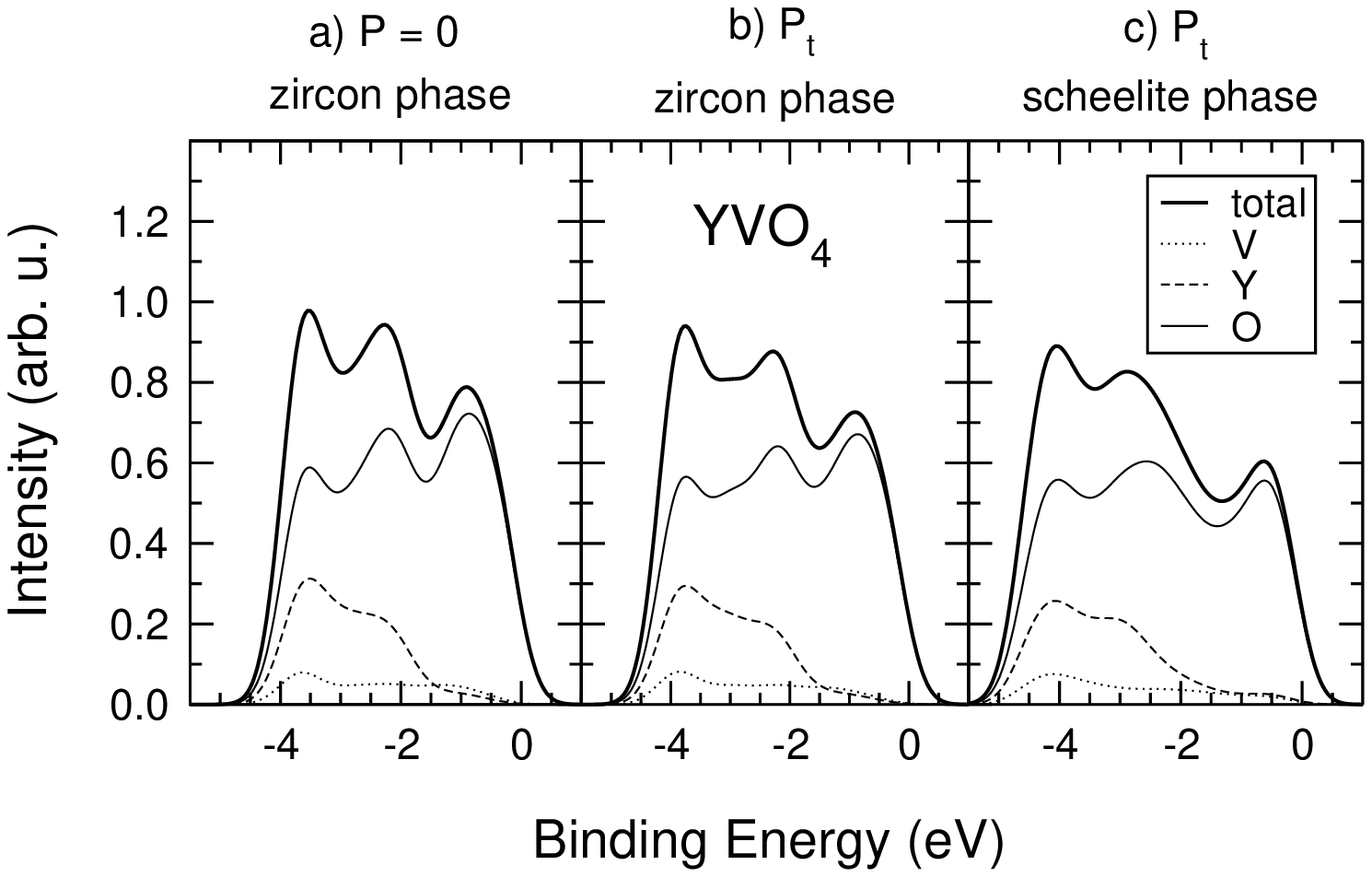}
\caption{\label{fig:xps} Theoretical X-Ray photoemission valence band spectra XPS for the zero pressure and calculated transition pressure P$_t$(GGA)~=~12.95~GPa for the zircon and scheelite-type phases of \yvo{}. Broadening parameter $\delta = 0.3$~eV.}
\end{figure}

\begin{figure}[ht]
\centering
\includegraphics[trim = 90 30 30 60,clip,height=\columnwidth,angle=270]{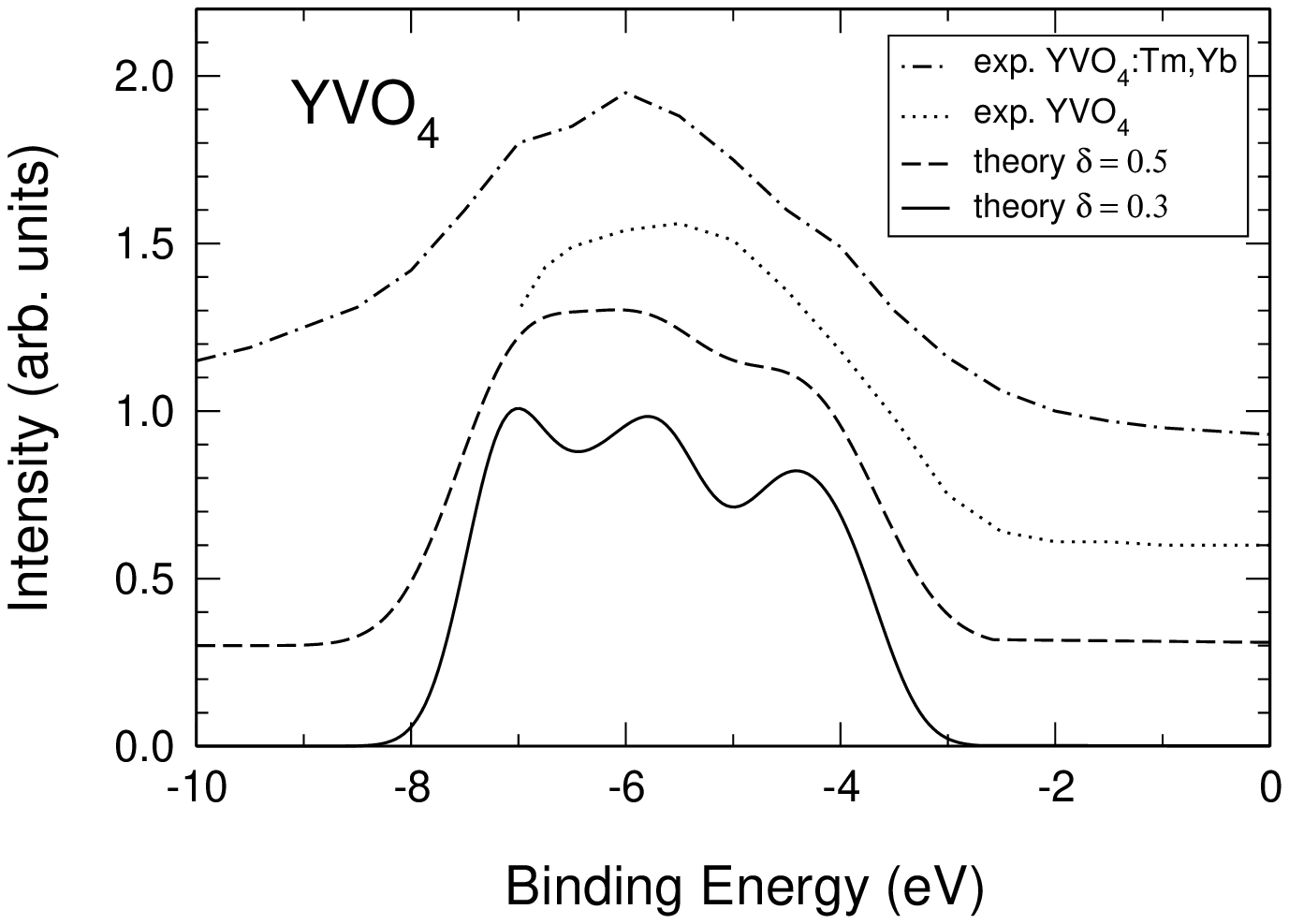}
\caption{\label{fig:xps_comp} 
Dash-dotted line for a sketch of experimental XPS for \yvo{}:Tm, Yb from available data~\cite{Kruczek05},
dotted line for a sketch of experimental XPS for \yvo{} from available data~\cite{Cai13},
dashed and solid lines for theoretical XPS for the zero pressure zircon-type phase of \yvo{} with broadening parameter $\delta$ equals 0.3 and 0.5~eV. 
}
\end{figure}

\subsection{Structure optimisation}

At ambient conditions \yvo{} crystallises in zircon-type structure (ZrSiO$_4$) and it transforms under pressure into scheelite-type phase (CaWO$_4$)~\cite{Wang04}, see Fig.~\ref{fig:structs}. 
The crystallographic structures of zircon and scheelite-type phases of \yvo{} have been fully optimised within the generalised gradient approximation (GGA-PBE)~\cite{Perdew96}.
The detailed results are compared with available experimental data~\cite{Wang04} and collected in Table~\ref{tab:crystal_data}.
The calculated lattice parameters at zero pressure are overestimated by about one percent which is attributed to the well known underbinding behaviour of the applied GGA.
On the same reason the calculated bulk moduli are most probably underestimated, what is not clear because of the limited accuracy of the experimental data~\cite{Wang04}.
The same underbinding behaviour of GGA for \yvo{} zircon and sheelite phases have been also observed previously~\cite{Manjon10, Huang12, Huang13}, see Table~\ref{tab:crystal_data}.
The relaxed atomic positions of oxygen atoms differ from the experiment on the third significant digit.

\section{Computational details}\label{sec:comp_details}

Scalar relativistic electronic band structure calculations are carried out by using Full-Potential Local-Orbital Minimum-Basis Scheme (FPLO-14)~\cite{Koepernik99} within the generalised gradient approximation (GGA) for the exchange-correlation potential in the Perdew, Burke, Ernzerhof form (PBE)~\cite{Perdew96}. 
Calculations are performed with $12 \times 12 \times 12$ \textbf{k}-mesh, 
energy convergence criterion $10^{-8}$~Ha and charge density convergence criterion $10^{-6}$.

Symmetry restricted geometry optimisations of internal parameters (Wyckoff positions) are performed with convergence criteria 10$^{8}$~Ha for the self-consistent electronic loop and 0.001 eV/\AA{} for the forces on each atom, 
see Table~\ref{tab:crystal_data}.
c/a ratios are optimised for every considered strained volume.
The equilibrium volumes, bulk moduli B$_0$, and transition pressure P$_t$ are obtained by fitting the third-order Birch-Murnaghan equation of state~\cite{Birch47} to the calculated energy-volume data.

The crystalline structure and electronic structure calculations are followed by a computational study of pressure dependent optical properties as electronic band gap and X-Ray photoemission spectra XPS.
Theoretical XPS valence band spectra are based directly on the band structure results. 
The partial densities of states are convoluted by the Gaussians with a full width at half maximum (FWHM) of $\delta = 0.3$~eV to mimic the experimental broadening coming from instrumental resolution, lifetime of hole states and thermal effects. 
The partial l--resolved densities of states DOS are multiplied by the corresponding photoemission cross--sections~\cite{Yeh85}. 
The accuracy of the partial photoemission cross--sections~\cite{Yeh85} is only two significant digits, 
thus the margin of error for the calculated intensities is about a few percent. 
The peak positions on the energy scale are as accurate as provided by GGA.
 
For visualisation of crystal structures VESTA code~\cite{Momma08} is used.

\section{Results and discussion}


In our study several structures with strained volumes are calculated for both zircon and scheelite phases in order to find energy-volume dependencies, see Fig.~\ref{fig:E_V}.
The full geometry optimisation is performed for every considered volume.
The energy-volume dependencies are then fit by the third-order Birch-Murnaghan equation of state~\cite{Birch47}.
The bulk moduli calculated from the third-order Birch-Murnaghan equation of state are 126~GPa and 142~GPa for zircon and scheelite-type phases, respectively.
They reasonably agree with experimental data (130(3)~GPa and 138(9)~GPa, respectively)~\cite{Wang04}.

The total energy vs. volume dependencies for the zircon and scheelite phases, as presented in Fig.~\ref{fig:E_V}, give the crossing point at V/V$_0$~=~0.919 indicating a cell volume of 304.198~\AA{}$^3$.
This point indicates the characteristic volume below which the scheelite phase become more energetically stable.
The transition pressure P$_t$ can be then read from the volume vs. pressure curve as obtained by fitting the third-order Birch-Murnaghan equation of state~\cite{Birch47} to the energy-volume data, see Fig.~\ref{fig:V_P}.
Hence, the theoretical P$_t$(GGA) equals 12.95~GPa which overestimates the experimental value $\sim$7.5~GPa~\cite{Panchal11}, what we attribute to the well known underbinding character of GGA.
Surprisingly, Manj\'{o}n et al.~\cite{Manjon10}, despite observing the same underbinding behaviour of GGA, are obtaining an underestimated value of zircon-scheelite phase transition P$_t$(GGA) equals 5.0~GPa.

c/a ratio and lattice parameters as a function of pressure are presented in Fig.~\ref{fig:lat_param_P}.
The theoretical pressure dependencies are in good qualitative agreement with the corresponding experimental values~\cite{Wang04}.

The atomic positions of oxygen atoms vary as a function of pressure for both zircon and scheelite phases, see Fig.~\ref{fig:positions_P}. 
It means that the tetrahedrons with oxygen atoms in vertices rotate in three dimensions, compare Fig.~\ref{fig:structs}.
The fact that tetrahedrons rotate with increase in pressure is one of the reasons why the full geometry should be optimised for energy-volume calculations, as we did in this work.
The comparison of oxygen positions to experimental data~\cite{Wang04} is unreliable because of the insufficient number of measured data points and their relatively large error bars.

\subsection{Electronic structure and optical properties}

\yvo{} and its derivatives attract a lot of attention mainly because its exceptional optical properties~\cite{Sato03,Ryba-Romanowski03,Huignard02,Mullica96}.
In this work we address theoretically some optical characteristic by studying the pressure dependencies of electronic band gaps and X-Ray photoemission spectra.

The calculated partial and total densities of states (DOS) for the zero pressure and for calculated transition pressure P$_t$(GGA)~=~12.95~GPa for the zircon and scheelite-type phases of \yvo{} have been presented in Fig.~\ref{fig:dos}. 
In all three cases \yvo{} exhibits an insulating character.

At theoretical P$_t$(GGA)~=~12.95~GPa the calculated within GGA band gaps are about 2.8~eV and 2.6~eV for zircon and scheelite phases, respectively (exp. 3.9~eV and 2.9~eV, respectively), see Fig.~\ref{fig:gap}. 
%
Besides the calculated band gaps are underestimated, also the increase of the band gap with pressure above the phase transition observed for scheelite phase disagrees with empirical data~\cite{Panchal11}.

The underestimation of the calculated band gaps indicates the insufficient description of the electronic correlations within GGA in \yvo{}.
This drawback has been improved by applying GGA+U scheme with the additional on-site Coulomb interactions on 3d orbitals of vanadium and 2p orbitals of oxygen. 
The reasonable value of Coulomb U for vanadium U(V)~=~3.5~eV have been taken from theoretical predictions for bulk vanadium~\cite{Sasioglu}.
The Coulomb U on oxygen U(O)~=~4.5~eV have been chosen close to the value taken by Korotin et al. for another oxide La$_{7/8}$Sr$_{1/8}$MnO$_3$~\cite{Korotin00}.
The band gap versus pressure dependency calculated within GGA+U method have been presented in Fig.~\ref{fig:gap}.
The application of correlations corrections elevates the band gaps values for both \yvo{} phases, what get the calculated results for zircon phase significantly closer to the experiment, but does not improve the agreement for scheelite phase.

%
%
%

The theoretical X-Ray photoemission valence band spectra XPS for the zero pressure and  P$_t$(GGA)~=~12.95~GPa for the zircon-type phase and for P$_t$(GGA) for  scheelite-type phase have been presented in Fig.~\ref{fig:xps}.
The calculated XPS are based on partial DOS as presented in Fig.~\ref{fig:dos}.
The XPS valence bands are comprised mainly of the oxygen contributions with smaller shares of yttrium and vanadium.
For the zircon phase at the transition pressure P$_t$ the broadening of the valence band is observed in comparison to the phase with zero pressure.

The comparison between the calculated XPS spectra and sketches of experimental results is presented in Fig.~\ref{fig:xps_comp}.
The experimental XPS data for a pure \yvo{}~\cite{Cai13} and for \yvo{}:Tm, Yb with about 2\% Tm and 5\% Yb substitutions on Y site~\cite{Kruczek05} have been taken from available data.
In XPS experiment the binding energy is often determined by reference to the C 1s line~\cite{Kruczek05}.
In the case of the materials with energy gap, like \yvo{}, it may lead to the non zero position of the edge of the valence band.
In theory, on the other hand, the Fermi level is determined based on the end of the valence band and the beginning of the band gap, which, by definition, places the edge of the valence band at the zero binding energy.
Thus, in order to compare experimental and theoretical spectra, the theoretical results have been shifted by arbitrary 3.5 eV, as presented in Fig.~\ref{fig:xps_comp}.
In the same figure the results for two broadening parameters $\delta$ equal 0.3 and 0.5 eV are presented.
It can be seen, that in the manner of a spectra resolution the overall similarity between theory and experiment is better for theoretical result with higher $\delta$, although the experimental result~\cite{Kruczek05} is labelled by authors with energy resolution 0.3 eV.
The valence band width of about 4~eV is in excellent agreement with experimental XPS result for doped \yvo{}~\cite{Kruczek05}.
The band width for the pure \yvo{} is most probably also close to 4~eV, but it can not be concluded with certainty because unfortunately the published spectrum~\cite{Cai13} does not cover the whole range of the valence band.

\section{Conclusions}

The parameter free \textit{ab inito} calculations have been performed in order to study the pressure induced structural phase transition and some optical properties of \yvo{}.
The full geometry have been optimised for all considered volumes.
The calculated zircon-scheelite transition pressure P$_t$(GGA) equals 12.95~GPa what still might be considered in reasonable agreement with experimental $\sim$7.5~GPa~\cite{Wang04}).
We attributed the discrepancy to underbinding character of the employed GGA.
All of the calculated structural pressure dependencies: V(P), a(P), c(P), and c/a(P) are in very good qualitative agreement with the experimental data~\cite{Wang04}.
The calculated pressure dependencies of Wyckoff positions for oxygen are difficult to compare to experimental results because of the insufficient number of the measured data points~\cite{Wang04}. 
The bulk moduli calculated from third-order Birch-Murnaghan equation of state equal 126~GPa (130(3)~GPa) and 142~GPa (138(9)~GPa) for zircon and scheelite-type phases respectively are in good agreement with empirical values (in parentheses)~\cite{Wang04}).
The calculated valence band width for zircon phase at zero pressure equals about 4~eV what is in very good agreement with the experimental XPS result  for doped \yvo{}~\cite{Kruczek05}.
The calculated band gaps at transition pressures P$_t$ equal 2.8~eV (3.9~eV) and 2.6~eV (2.9~eV) for zircon and scheelite phases respectively and underestimate the experimental values (in parentheses)~\cite{Panchal11}.
The application of GGA+U significantly improves the band gaps for zircon phase.

\section{Acknowledgements}

Work supported by the National Science Centre (Poland) grant: DEC-2011/01/B/ST3/02212 .

\end{sloppypar}

\end{document}